\documentclass[12pt]{article}
\usepackage{amsfonts}
\usepackage{amsmath, amssymb}
\usepackage{youngtab}
\usepackage{hyperref}
\usepackage[pdftex]{graphicx}
\usepackage{braket}
\usepackage{bm}
\usepackage{subcaption}

\newcommand{\I}{\mathcal I}
\newcommand{\cH}{\mathcal H}
\newcommand{\Ftwo}{\mathbb F_2}

\usepackage{comment}
\includecomment{pdffig}

\allowdisplaybreaks[1]

\Yboxdim{5pt}

\makeatletter
    
    \@addtoreset{equation}{section}
  \makeatother

\usepackage{color}

\def\rem#1{}

\renewcommand{\title}[1]{\vbox{\center\LARGE{#1}}\vspace{5mm}}
\renewcommand{\author}[1]{\vbox{\center\large#1}\vspace{5mm}}

\begin{document}

\begin{titlepage}
\begin{center}
\vspace{5mm}
\hfill {\tt 
}\\
\vspace{20mm}

\title{
\LARGE
Topological Modular Forms Constrain Threshold States in Extremal $\mathcal N=1$ AdS$_3$ Gravity
}
\vspace{7mm}

Yutaka Yoshida

\vspace{6mm}

\vspace{3mm}
{\small {\it Department of Current Legal Studies, Faculty of Law, Meiji Gakuin University, 1-2-37 
Shirokanedai, Minato-ku, Tokyo 108-8636, Japan}} \\
{\small {\it Institute for Mathematical Informatics, Meiji Gakuin University, 
1518 Kamikurata-cho, Totsuka-ku, Yokohama 244-8539, Japan}}
 \\

\end{center}

\vspace{7mm}
\begin{abstract}
We identify a topological constraint on extremal $\mathcal N=1$ AdS$_3$ spectra that is not implied by genus-one modular covariance of the three even spin structure partition functions.
Assuming the expected topological modular forms divisibility of the Ramond Witten index, we show that for infinitely many odd $n$ at central charge $c=12n$, an extremal holomorphic $\mathcal N=1$ superconformal field theory requires an odd number of Neveu--Schwarz primaries at the BTZ threshold.
Thus the strict ansatz with no threshold primaries is excluded for this infinite family.
\end{abstract}
\vfill

\end{titlepage}


\section{Introduction}
Three-dimensional gravity with negative cosmological constant provides a particularly constrained setting for holography, owing to the powerful modular properties of the dual two-dimensional conformal field theory. Its asymptotic symmetry algebra contains two Virasoro algebras with the Brown--Henneaux central charge~\cite{Brown:1986nw}, while the massless BTZ black hole marks a natural threshold in the boundary spectrum~\cite{Banados:1992wn}. In a holomorphically factorized description, Witten proposed extremal chiral conformal field theories as candidate duals of pure gravity and pure supergravity~\cite{Witten:2007kt}. 
Here we focus on the extremal holomorphic $\mathcal N=1$ superconformal field theories (SCFTs) proposed there as candidate holographic duals of minimal AdS$_3$ supergravity. Their spectrum below the BTZ threshold consists only of states in the vacuum module of the $\mathcal N=1$ superconformal algebra. The holomorphic central charge is
\begin{align}
c=12n,\qquad n=1,2,3,\ldots\,.
\end{align}
Witten's extremality condition allows Neveu--Schwarz (NS) primaries exactly at
\begin{align}
h=\frac{c}{24}=\frac n2\,.
\end{align}
We denote the number of such threshold primaries by $s$ and refer to $s=0$ as the \emph{strict} extremal ansatz.

Our additional input comes from topological modular forms (TMF)~\cite{Hopkins:2002}. 
The Stolz--Teichner conjecture relates two-dimensional $\mathcal N=(0,1)$ supersymmetric field theories to classes in TMF~\cite{Stolz:2011zj}.
 For a holomorphic $\mathcal N=1$ SCFT with $c=12n$, let
$\I:=\operatorname{Tr}_{\cH_{\rm R}}(-1)^F$ denote the Ramond Witten index. 
The corresponding TMF class maps under the Witten genus to the modular form $\I\Delta^n$, where $\Delta=\eta^{24}$ is the modular discriminant~\cite{Hopkins:2002, Gaiotto:2018ypj, Tachikawa:2021mvw}.
Not every integral modular form lies in the image of the Witten genus.
In particular, the smallest positive multiple of \(\Delta^n\) in the image 
is $\frac{24}{\gcd(24,n)}\Delta^n$~\cite{Hopkins:2002,Gaiotto:2018ypj,Tachikawa:2021mvw}. The expected relation between SCFTs and TMF therefore predicts
\begin{align}
 \frac{24}{\gcd(24,n)}\mid \I.
 \label{eq:tmfdiv}
\end{align}
The identification of a general interacting holomorphic $\mathcal N=1$ SCFT with a TMF class remains conjectural, although the predicted divisibility has been verified in nontrivial classes of examples~\cite{Gaiotto:2018ypj,Albert:2022gcs}.

Assuming Eq.~\eqref{eq:tmfdiv}, we show that for infinitely many odd $n$, the number of ${\rm NS}$ primaries at the BTZ threshold must be odd:
\begin{align}
s\equiv1\pmod{2}.
 \label{eq:intro-main}
\end{align}
Thus the strict extremal ansatz is excluded for an infinite family. 
In Witten's construction, the ${\rm NS}$ extremal data determine a quantity $\beta_n$ through modular covariance. For a general extremal spectrum with $s$ ${\rm NS}$ primaries at the threshold,
\begin{align}
\beta_n
=
h_0-(-1)^n s.
\label{eq:h0threshold}
\end{align}
Here $h_0$ is the total number of Ramond ground states; the modular data determine $\beta_n$, but not $h_0$ and $s$ separately.
The TMF divisibility supplies an additional parity constraint. Since the Witten index satisfies $\I\equiv h_0\pmod 2$, the required divisibility of $\I$ constrains the parity of $h_0$ and hence of $s$. Our arithmetic result shows that $\beta_n$ is odd for infinitely many odd $n$, and therefore forces $s$ to be odd for this infinite family. The arithmetic results do not rely on the conjectural relation between SCFTs and TMF; TMF enters only through Eq.~\eqref{eq:tmfdiv}.

Previous tests of the extremal program include the gravitational partition function and holomorphic factorization~\cite{Maloney:2007ud}, modular differential equations and $C_2$-cofiniteness~\cite{Gaberdiel:2007ve}, higher-genus consistency~\cite{Gaberdiel:2010jf}, modular bootstrap~\cite{Bae:2016yna,Bae:2018qym}, and bounds for self-dual vertex operator superalgebras~\cite{Hoehn2008}. Ramond sector constraints and the role of spin structure partition functions were emphasized in~\cite{Benjamin:2020zbs}, while TMF constraints on bosonic holomorphic extremal CFTs were studied in~\cite{Lin:2021bcp}. The mechanism considered here is different from near-extremal modifications motivated by small black holes~\cite{Benjamin:2016aww}: through the Ramond Witten index, TMF constrains the parity of the number of ${\rm NS}$ primaries exactly at the BTZ threshold.

\section{Extremal spectra and the \texorpdfstring{$c=36$}{c=36} example}
The distinction between the genus-one modular constraints and the TMF divisibility condition 
can be seen directly from the four torus spin structures. 
With $q=e^{2\pi i\tau}$, we adopt the convention of~\cite{Witten:2007kt} that absorbs the Casimir shift into $L_0$. The holomorphic torus amplitudes are
\begin{align}
Z_{{\rm NS}+}&=\operatorname{Tr}_{\cH_{\rm NS}}q^{L_0},
\qquad
Z_{{\rm NS}-}=\operatorname{Tr}_{\cH_{\rm NS}}(-1)^Fq^{L_0},
\nonumber\\
Z_{{\rm R}+}&=\operatorname{Tr}_{\cH_{\rm R}}q^{L_0},
\qquad
Z_{{\rm R}-}=\operatorname{Tr}_{\cH_{\rm R}}(-1)^F=\I.
\end{align}
The three even spin structures are permuted among themselves by modular transformations, 
whereas the partition function for the odd spin structure,
$Z_{{\rm R}-}$, is the $q$-independent Witten index and is not determined by the three even spin structures~\cite{Witten:2007kt,Benjamin:2020zbs}. This distinction is especially transparent in the Ramond sector. 
In the Ramond sector, the supercharge zero mode obeys
\begin{align}
G_0^2=L_0.
\end{align}
Every state with $L_0>0$ is therefore paired with a state of opposite fermion parity. Such pairs contribute twice to $Z_{{\rm R}+}$ and cancel from $Z_{{\rm R}-}$. On Ramond ground states, $L_0=0$, so $G_0$ acts trivially in a unitary theory, and there is no analogous pairing constraint. Thus the constant term of $Z_{{\rm R}+}$ gives the total number of Ramond ground states, whereas $Z_{{\rm R}-}=\I$ gives the difference between the numbers of bosonic and fermionic Ramond ground states. Genus-one modular covariance constrains the former through the three even spin structures, but does not determine the latter. The TMF divisibility instead constrains the Witten index.

At $c=36$ ($n=3$), the effect is already explicit.
Let $K(\tau)$ denote the modular function defined in Appendix~\ref{app:parity-proof}.
The Ramond cusp may be represented by $\tau=1$, where $K(1)=-24$.
Let $Z_{{\rm NS}+}^{\rm strict}(\tau)$ denote the ${\rm NS}$ partition function in the strict extremal ansatz $s=0$.
It can be written as
\begin{align}
Z_{{\rm NS}+}^{\rm strict}(\tau)
=
F_3\bigl(K(\tau)\bigr),
\end{align}
where~\cite{Witten:2007kt}
\begin{align}
F_3(x)=x^3-828x-6143.
\end{align}
Applying the modular transformation to the Ramond spin structure gives
\begin{align}
Z_{{\rm R}+}^{\rm strict}(\tau)
&=
-
Z_{{\rm NS}+}^{\rm strict}\left(1-\frac{1}{\tau}\right)
\nonumber\\
&=
95+3686400q+1296433152q^2+\cdots.
\end{align}
Thus the number of Ramond ground states in the strict ansatz is
\begin{align}
h_0^{\rm strict}(3)
=
-Z_{{\rm NS}+}^{\rm strict}(1)
=
-F_3\bigl(K(1)\bigr)
=
95.
\end{align}
Adding threshold primaries has a particularly simple modular effect. An ${\rm NS}$ primary at $h=c/24$ has $L_0=0$, so each such primary adds one to the constant term of $Z_{{\rm NS}+}$ without changing its polar part. Under the modular transformation to the Ramond channel, a constant shift $s$ contributes $(-1)^n s$ to the constant term of $Z_{{\rm R}+}$~\cite{Witten:2007kt}. Since $n=3$, allowing $s$ such ${\rm NS}$ primaries changes the total number of Ramond ground states to
\begin{align}
h_0=
N_B+N_F=\beta_3-s=
95-s,
\label{eq:h036}
\end{align}
where $N_B$ and $N_F$ denote the numbers of bosonic and fermionic Ramond ground states, respectively.
Their difference is the Witten index,
\begin{align}
 \I=N_B-N_F\equiv N_B+N_F=h_0\pmod{2}.
 \label{eq:Iparity}
\end{align}
For $n=3$, Eq.~\eqref{eq:tmfdiv} requires $8\mid\I$, hence $\I$ is even. Equation~\eqref{eq:h036} therefore implies Eq.~\eqref{eq:intro-main} in this example.

The even spin torus data do not determine the Witten index itself. In the strict $c=36$ spectrum, however, they determine its parity: since $h_0=95$, any decomposition $95=N_B+N_F$ gives $\I=N_B-N_F\equiv95\equiv1\pmod2$. The expected TMF divisibility instead requires $8\mid\I$. Allowing threshold primaries changes $h_0$ to $95-s$, and compatibility with TMF therefore requires $s$ to be odd. Thus the $c=36$ example shows explicitly how the Witten index constrains the number of ${\rm NS}$ primaries at the threshold.

For general $n$, the ${\rm NS}$ partition function in the strict extremal ansatz can similarly be written as
\begin{align}
Z_{{\rm NS}+}^{\rm strict}(\tau)=F_n\bigl(K(\tau)\bigr),
\end{align}
where $F_n$ is a polynomial in $K$ uniquely fixed by matching the polar part of the ${\rm NS}$ vacuum character.
The quantity $\beta_n$ introduced above is then given by
\begin{align}
\beta_n=(-1)^nF_n(-24).
\label{eq:betadef}
\end{align}

\section{Parity theorem for an infinite family}
We first express the parity of $\beta_n$ in terms of the coefficients of the $\mathcal N=1$ vacuum character and then show that these coefficients are odd for infinitely many odd $n$. 
Set $t=q^{1/2}$ and define the descendant contribution to the ${\rm NS}$ vacuum character by
\begin{align}
D(t):=\prod_{m=2}^{\infty}\frac{1+t^{2m-1}}{1-t^{2m}}
=\sum_{r\geq0}d_r t^r.
\end{align}
Here the numerator arises from fermionic descendants generated by the superconformal modes $G_{-(m-1/2)}$, while the denominator comes from bosonic descendants generated by the Virasoro modes $L_{-m}$. The product starts at $m=2$ because $G_{-1/2}$ and $L_{-1}$ annihilate the vacuum.
With this notation, the matching condition that fixes $F_n$ takes the form
\begin{align}
F_n\bigl(K(\tau)\bigr)=t^{-n}D(t)+O(t).
\label{eq:extremalmatching}
\end{align}
The relevant parity relation is
\begin{equation}
\beta_n\equiv d_n\pmod{2}.
\label{eq:betaparity}
\end{equation}
Its derivation is given in Appendix~\ref{app:parity-proof}. Modulo two, every positive power of $K(\tau)$, regarded as a Laurent series in $t$, has vanishing constant term; taking the constant term of Eq.~\eqref{eq:extremalmatching} therefore reduces the Ramond parity problem to a single coefficient of the universal product $D(t)$.

For infinitely many odd $n$, $\beta_n$ is odd:
\begin{equation}
 \beta_n \equiv1\pmod2
 \quad\text{for infinitely many odd }n.
 \label{eq:infiniteodd}
\end{equation}
To prove this, write $E(t)=\prod_{m\ge1}(1-t^m)$. In characteristic two,
\begin{align}
D(t)\equiv\frac{1+t}{E(t)^3},
 \qquad
 E(t)^3\equiv\sum_{j\ge0}t^{j(j+1)/2}.
\end{align}
The contradiction comes from the exponents that can appear on the two sides. 
If only finitely many coefficients $d_n$ with odd $n$ were odd, differentiation of the identity $D(t)E(t)^3=1+t$ would give, after setting $u=t^2$,
\begin{align}
R(u)\sum_{j\ge0}u^{j(j+1)/2}
=
\sum_{r\ge2}u^{\lfloor r(r+1)/4\rfloor}
\end{align}
for some polynomial $R$. The exponents on the left lie in intervals of fixed width above the triangular numbers. The gaps between successive triangular numbers grow without bound, whereas the exponents $\lfloor r(r+1)/4\rfloor$ on the right enter those gaps. This contradiction proves Eq.~\eqref{eq:infiniteodd}; the complete argument is given in Appendix~\ref{app:parity-proof}.

For example, Witten's values of $\beta_n$ for the first few odd $n$ are~\cite{Witten:2007kt}
\[
\begin{array}{c|ccccc}
 n & 1 & 3 & 5 & 7 & 9\\ \hline
 \beta_n & 24 & 95 & 143 & 262 & 453\\
 \text{TMF constraint on }s\!\!\pmod 2 & 0 & 1 & 1 & 0 & 1
\end{array}
\]

\section{TMF and threshold states}
For odd $n$,
\begin{align}
 \frac{24}{\gcd(24,n)}\in\{8,24\},
\end{align}
so the expected TMF divisibility makes $\I$ even. Combining Eqs.~\eqref{eq:Iparity} and \eqref{eq:h0threshold}, one has for every odd $n$
\begin{equation}
 \I\equiv h_0 \equiv \beta_n - s\pmod{2}.
 \label{eq:generalparity}
\end{equation}
Hence, for every odd $n$ in the infinite family~\eqref{eq:infiniteodd}, the oddness of $\beta_n$ together with the evenness of $\I$ requires $s$ to be odd. Assuming the expected TMF divisibility, there are therefore infinitely many odd $n$ for which 
 every extremal holomorphic $\mathcal N=1$ SCFT at $c=12n$ must contain an odd number of ${\rm NS}$ primaries exactly at the BTZ threshold. The strict ansatz $s=0$ is excluded throughout this infinite family.

Equation~\eqref{eq:generalparity} fixes only the parity of the threshold correction. In particular, Eq.~\eqref{eq:intro-main} is a necessary condition, not an existence theorem for an extremal SCFT with odd $s$. 
Nor does the full divisibility condition in Eq.~\eqref{eq:tmfdiv} determine $s$ modulo $8$ or $24$, because the Witten index depends on the difference between the numbers of bosonic and fermionic Ramond ground states, not only on their total number.

\section{Discussion}
Equation~\eqref{eq:intro-main} has a natural bulk interpretation. The states counted by $s$ are additional ${\rm NS}$ primaries, not states in the vacuum module of the $\mathcal N=1$ superconformal algebra, and they sit precisely at $h=c/24$, the BTZ threshold in the holomorphically factorized extremal construction~\cite{Witten:2007kt}. They are therefore absent from the perturbative supergravity vacuum module and lie exactly at the threshold. For the infinite family identified above, compatibility with the expected TMF divisibility requires an odd number of such threshold primaries; in particular, the threshold sector cannot be empty.

This constraint is complementary to other tests of extremal theories. Our argument uses the Ramond Witten index, whose value is not fixed by the modular orbit of the three even spin structures~\cite{Witten:2007kt,Benjamin:2020zbs}. Its parity equals that of the total number of Ramond ground states. At $c=36$, for example, the strict extremal ansatz gives $95$ Ramond ground states and hence forces the Witten index to be odd, in conflict with the TMF requirement $8\mid\I$. Thus TMF supplies information not contained in the even spin torus data.

The argument also differs from the bosonic TMF obstruction of~\cite{Lin:2021bcp}. Here the ${\rm NS}$ extremal data determine the total number of Ramond ground states only up to the threshold contribution in Eq.~\eqref{eq:h0threshold}, while the Witten index measures the difference between the numbers of bosonic and fermionic Ramond ground states. Combining these two pieces of information converts the TMF divisibility into the parity constraint~\eqref{eq:intro-main}, which is intrinsically $\mathcal N=1$.

Equation~\eqref{eq:tmfdiv} is an expected consequence of the Stolz--Teichner conjecture rather than a theorem for arbitrary interacting holomorphic $\mathcal N=1$ SCFTs, so the AdS$_3$ constraint is conditional on the conjectural SCFT--TMF correspondence. It would be interesting to understand the bulk origin of the threshold states required by this constraint.

\appendix
\section{Proof of the parity statements}\label{app:parity-proof}

We prove Eqs.~\eqref{eq:betaparity} and \eqref{eq:infiniteodd}. Viewing $K(\tau)$ as 
a Laurent series in $t$, the modular function satisfies
\begin{align}
K(\tau)+24
 =t^{-1}\frac{E(t^2)^{48}}{E(t^4)^{24}E(t)^{24}}.
\end{align}
Working modulo two with formal Laurent series, i.e. in $\Ftwo((t))$, the Frobenius identity $f(t)^{2^k}=f(t^{2^k})$ implies
\begin{align}
E(t^2)^{48}=E(t^{32})^3,
 \qquad
 E(t^4)^{24}=E(t^{32})^3,
\end{align}
and hence
\begin{align}
K(\tau)\equiv t^{-1}E(t^8)^{-3}.
\end{align}
For every $r>0$, write $r=2^v m$ with $m$ odd. Then
\begin{align}
K(\tau)^r\equiv t^{-r}E(t^{2^{v+3}})^{-3m}.
\end{align}
Its exponents have the form $-r+2^{v+3}\ell$ with $\ell\ge0$. Such an exponent cannot vanish because $r=2^v m$ with $m$ odd. Hence the constant term of $K(\tau)^r$ vanishes modulo two for every $r>0$. Since $K(\tau)=t^{-1}+O(t)$ and both $K(\tau)$ and $D(t)$ have integral coefficients, the matching condition~\eqref{eq:extremalmatching} determines $F_n$ recursively with integer coefficients. Writing $F_n(x)=\sum_{r=0}^{n}f_{n,r}x^r$ with $f_{n,r}\in\mathbb Z$, it follows that the constant term of $F_n(K(\tau))$ is congruent to $f_{n,0}$ modulo two. Since $(-1)^n\equiv1\pmod2$ and $(-24)^r$ vanishes modulo two for $r>0$, Eq.~\eqref{eq:betadef} gives $\beta_n \equiv f_{n,0}\pmod{2}$. Taking the constant term of Eq.~\eqref{eq:extremalmatching} gives $f_{n,0}\equiv d_n\pmod{2}$, proving Eq.~\eqref{eq:betaparity}.

It remains to prove Eq.~\eqref{eq:infiniteodd}. Reduce all series in the remainder of the proof modulo two. From the definition of $D(t)$,
\begin{align}
D(t)
&\equiv
\frac{\prod_{m\geq2}(1+t^{2m-1})}
     {\prod_{m\geq2}(1+t^{2m})}
\nonumber\\
&=
\frac{(1+t)E(t)}{E(t^2)^2}
=
\frac{1+t}{E(t)^3}
\pmod{2},
\end{align}
where we used
\begin{align}
E(t)
=
\prod_{m\geq1}(1+t^{2m-1})E(t^2)
\pmod{2}
\end{align}
and $E(t^2)=E(t)^2$ over $\Ftwo$.
A standard specialization of Jacobi's triple product identity,
\begin{align}
E(t)^3=\sum_{j\ge0}(-1)^j(2j+1)t^{j(j+1)/2},
\end{align}
reduces modulo two to
\begin{align}
T(t):=E(t)^3
 \equiv\sum_{j\ge0}t^{\nu_j},
 \qquad
 \nu_j=\frac{j(j+1)}2,
\end{align}
and hence
\begin{equation}
 D(t)T(t)=1+t.
 \label{eq:DTEEnd}
\end{equation}
Assume that only finitely many odd $n$ have $d_n\equiv1\pmod{2}$. Differentiating in characteristic two kills every even power and sends each surviving odd power to an even one, so
\begin{align}
D'(t)=R(t^2)
\end{align}
for some polynomial $R\in\Ftwo[u]$. If $R=0$, differentiating Eq.~\eqref{eq:DTEEnd} and using Eq.~\eqref{eq:DTEEnd} again gives $T(t)+(1+t)T'(t)=0$, contradicting the explicit nonzero series in Eq.~\eqref{eq:TderivativeEnd} below. We may therefore assume $R\ne0$.
Differentiating Eq.~\eqref{eq:DTEEnd} and multiplying by $T$ yields
\begin{align}
D'(t)T(t)^2=T(t)+(1+t)T'(t).
\end{align}
By the same Frobenius identity, $T(t)^2=T(t^2)$. For $j\ge2$, the contribution of the monomial $t^{\nu_j}$ to $T+(1+t)T'$ is $t^{\nu_j}$ when $\nu_j$ is even and $t^{\nu_j-1}$ when $\nu_j$ is odd: in the latter case the two terms at degree $\nu_j$ cancel in characteristic two. Hence its exponent is $2\lfloor\nu_j/2\rfloor$. Set
\begin{align}
 a_j:=\left\lfloor\frac{j(j+1)}4\right\rfloor.
\end{align}
Since $((j+1)(j+2)-j(j+1))/4=(j+1)/2>1$ for $j\ge2$, the integers $a_j$ are strictly increasing, so these exponents are distinct. The combined contributions from $j=0$ and $j=1$ cancel. Thus
\begin{equation}
 T(t)+(1+t)T'(t)
 =\sum_{r\ge2}t^{2a_r}.
 \label{eq:TderivativeEnd}
\end{equation}
Setting $u=t^2$ gives
\begin{equation}
 R(u)\sum_{j\ge0}u^{\nu_j}
 =\sum_{r\ge2}u^{a_r}.
 \label{eq:RTSEnd}
\end{equation}
Write $R(u)=\sum_{d=0}^{M}r_d u^d$, where $M=\deg R$. Every exponent appearing on the left-hand side of Eq.~\eqref{eq:RTSEnd} has the form $\nu_j+d$ with $r_d\ne0$ and $0\le d\le M$. 
Hence these exponents lie in the intervals $[\nu_j,\nu_j+M]$. Whenever $k$ is large enough that $k+1>M$, no exponent on the left lies in the open interval $(\nu_k+M,\nu_{k+1})$.

We now show that the right-hand side must enter one of these gaps. Set
$N_k=\lceil3k/2\rceil$. Since
\begin{align}
a_{N_k}\ge \frac{9k^2}{16}+\frac{3k}{8}-1,
\end{align}
we have $a_{N_k}>\nu_k+M$ for all sufficiently large $k$. Let $r$ be minimal with $a_r>\nu_k+M$. Then $r\le N_k$, while minimality and the elementary bound
\begin{align}
a_r-a_{r-1}\le\left\lceil\frac r2\right\rceil,
\end{align}
which follows from $\lfloor x\rfloor-\lfloor y\rfloor\le\lceil x-y\rceil$, imply
\begin{align}
a_r\le\nu_k+M+\left\lceil\frac{N_k}{2}\right\rceil.
\end{align}
Taking $k$ sufficiently large so that also
$M+\lceil N_k/2\rceil<k+1=\nu_{k+1}-\nu_k$, we obtain
\begin{align}
\nu_k+M<a_r<\nu_{k+1}.
\end{align}
The monomial $u^{a_r}$ therefore occurs with coefficient one on the right-hand side of Eq.~\eqref{eq:RTSEnd} but lies in a gap of the left-hand side, a contradiction. Hence $d_n\equiv1\pmod{2}$ for infinitely many odd $n$, and Eq.~\eqref{eq:betaparity} proves Eq.~\eqref{eq:infiniteodd}.


\begin{thebibliography}{99}

\bibitem{Brown:1986nw}
J.~D.~Brown and M.~Henneaux,
``Central Charges in the Canonical Realization of Asymptotic Symmetries: An Example from Three-Dimensional Gravity,''
Commun. Math. Phys. \textbf{104}, 207-226 (1986)
doi:10.1007/BF01211590

\bibitem{Banados:1992wn}
M.~Banados, C.~Teitelboim and J.~Zanelli,
``The Black hole in three-dimensional space-time,''
Phys. Rev. Lett. \textbf{69}, 1849-1851 (1992)
doi:10.1103/PhysRevLett.69.1849
[arXiv:hep-th/9204099 [hep-th]].

\bibitem{Witten:2007kt}
E.~Witten,
``Three-Dimensional Gravity Revisited,''
[arXiv:0706.3359 [hep-th]].


\bibitem{Hopkins:2002}
M.~J.~Hopkins,
``Algebraic topology and modular forms,''
in \textit{Proceedings of the International Congress of Mathematicians, Beijing 2002},
Vol.~I, pp.~291--317,
Higher Education Press, Beijing (2002)
[arXiv:math/0212397 [math.AT]].


\bibitem{Stolz:2011zj}
S.~Stolz and P.~Teichner,
``Supersymmetric field theories and generalized cohomology,''
in \emph{Mathematical Foundations of Quantum Field Theory and Perturbative String Theory},
Proc. Symp. Pure Math. \textbf{83}, 279 (2011)
[arXiv:1108.0189 [math.AT]].

\bibitem{Gaiotto:2018ypj}
D.~Gaiotto and T.~Johnson-Freyd,
``Holomorphic SCFTs with small index,''
Can. J. Math. \textbf{74}, no.2, 573-601 (2022)
doi:10.4153/S0008414X2100002X
[arXiv:1811.00589 [hep-th]].

\bibitem{Tachikawa:2021mvw}
Y.~Tachikawa,
``Topological modular forms and the absence of a heterotic global anomaly,''
PTEP \textbf{2022} (2022) no.4, 04A107
doi:10.1093/ptep/ptab060
[arXiv:2103.12211 [hep-th]].


\bibitem{Albert:2022gcs}
J.~Albert, J.~Kaidi and Y.~H.~Lin,
``Topological modularity of supermoonshine,''
PTEP \textbf{2023} (2023) no.3, 033B06
doi:10.1093/ptep/ptad034
[arXiv:2210.14923 [hep-th]].

\bibitem{Maloney:2007ud}
A.~Maloney and E.~Witten,
``Quantum Gravity Partition Functions in Three Dimensions,''
JHEP \textbf{02}, 029 (2010)
doi:10.1007/JHEP02(2010)029
[arXiv:0712.0155 [hep-th]].


\bibitem{Gaberdiel:2007ve}
M.~R.~Gaberdiel,
``Constraints on extremal self-dual CFTs,''
JHEP \textbf{11} (2007), 087
doi:10.1088/1126-6708/2007/11/087
[arXiv:0707.4073 [hep-th]].


\bibitem{Gaberdiel:2010jf}
M.~R.~Gaberdiel, C.~A.~Keller and R.~Volpato,
``Genus Two Partition Functions of Chiral Conformal Field Theories,''
Commun. Num. Theor. Phys. \textbf{4} (2010), 295-364
doi:10.4310/CNTP.2010.v4.n2.a2
[arXiv:1002.3371 [hep-th]].


\bibitem{Bae:2016yna}
J.~B.~Bae, K.~Lee and S.~Lee,
``Bootstrapping Pure Quantum Gravity in AdS3,''
[arXiv:1610.05814 [hep-th]].

\bibitem{Bae:2018qym}
J.~B.~Bae, S.~Lee and J.~Song,
``Modular constraints on superconformal field theories,''
JHEP \textbf{01} (2019), 209
doi:10.1007/JHEP01(2019)209
[arXiv:1811.00976 [hep-th]].

\bibitem{Hoehn2008}
G.~H\"ohn,
``Self-Dual Vertex Operator Superalgebras of Large Minimal Weight,''
arXiv:0801.1822 [math.QA].

\bibitem{Benjamin:2020zbs}
N.~Benjamin and Y.~H.~Lin,
``Lessons from the Ramond sector,''
SciPost Phys. \textbf{9} (2020) no.5, 065
doi:10.21468/SciPostPhys.9.5.065
[arXiv:2005.02394 [hep-th]].

\bibitem{Lin:2021bcp}
Y.~H.~Lin and D.~Pei,
``Holomorphic CFTs and Topological Modular Forms,''
Commun. Math. Phys. \textbf{401} (2023) no.1, 325-332
doi:10.1007/s00220-023-04639-3
[arXiv:2112.10724 [hep-th]].


\bibitem{Benjamin:2016aww}
N.~Benjamin, E.~Dyer, A.~L.~Fitzpatrick, A.~Maloney and E.~Perlmutter,
``Small Black Holes and Near-Extremal CFTs,''
JHEP \textbf{08} (2016), 023
doi:10.1007/JHEP08(2016)023
[arXiv:1603.08524 [hep-th]].




\end{thebibliography}

\end{document}